\documentclass{tlp}

\usepackage{latexsym}
\usepackage{psfig}

\newcommand\bcmdtab{\noindent\bgroup\tabcolsep=0pt%
  \begin{tabular}{@{}p{10pc}@{}p{20pc}@{}}}
\newcommand\ecmdtab{\end{tabular}\egroup}

\newcommand{\At}{\mbox{{\it At}}}
\newcommand{\n}{\mbox{{\bf not}}}
\newcommand{\lfp}{\mbox{{\it lfp}}}
\newcommand{\res}{\mbox{{\it res}}}
\newcommand{\false}{\mbox{{\it false}}}
\newcommand{\findset}{\mbox{{\it findset}}}
\newcommand{\pred}{\mbox{{\it pred}}}
\newcommand{\wfs}{{\small\it wfs}}

\newtheorem{theorem}{Theorem}[section]
\newtheorem{lemma}[theorem]{Lemma}

  \title[Theory and Practice of Logic Programming]
        {On the problem of computing the well-founded
         semantics\footnote{A preliminary version of this paper 
         appeared in the Proceedings of Computational Logic -- CL 2000,
         Lecture Notes in Artificial Intelligence, 1861, Springer
         Verlag, 2000. }}

  \author[Z. Lonc and M. Truszczy\'nski]
         {Zbigniew Lonc\thanks{On leave from Warsaw University of
         Technology.} and Miros\l aw Truszczy\'nski\\
         Department of Computer Science, University of Kentucky,
         Lexington, KY 40506-0046, USA\\
         \email{lonc|mirek@cs.uky.edu}}

\pagerange{\pageref{firstpage}--\pageref{lastpage}}

\begin{document}

\label{firstpage}

\maketitle

\begin{abstract}
The well-founded semantics is one of the most widely studied and used
semantics of logic programs with negation. In the case of finite
propositional programs, it can be computed in polynomial time, more
specifically, in $O(|\At(P)|\times size(P))$ steps, where $size(P)$
denotes the total number of occurrences of atoms in a logic program $P$.
This bound is achieved by an algorithm introduced by Van Gelder and
known as the alternating-fixpoint algorithm. Improving on the 
alternating-fixpoint algorithm turned out to be difficult. In this 
paper we study extensions and modifications of the alternating-fixpoint
approach. We then restrict our attention to the class of programs whose
rules have no more than one positive occurrence of an atom in their
bodies. For programs in that class we propose a new implementation of
the alternating-fixpoint method in which false atoms are computed in a
top-down fashion. We show that our algorithm is faster than other
known algorithms and that for a wide class of programs it is linear and
so, asymptotically optimal.
\end{abstract}

\section{Introduction}
\label{intro}

The well-founded semantics was introduced in \cite{vrs91} to provide 
3-valued interpretations to logic programs with negation. Since its 
introduction, the well-founded semantics has become one of the most 
widely studied and most commonly accepted approaches to negation in 
logic programming \cite{adp95,fi91a,csw94,cw96,zbf97,bd98}. It was
implemented in several top-down reasoning systems, most prominent of
which is XSB \cite{ssw97}. 

The well-founded semantics is closely related to the stable-model semantics 
\cite{gl88}, another major approach to logic programs with negation. 
The well-founded semantics approximates the stable-model semantics 
\cite{vrs91,fi99}. Moreover, computing the well-founded model of 
propositional programs is polynomial \cite{vg89} while computing stable 
models is NP-hard \cite{mt88}. Consequently, evaluating the well-founded 
semantics can be used as an effective preprocessing technique in 
algorithms to compute stable models \cite{vsnv95}. In addition, as 
demonstrated by {\tt smodels} \cite{ns96}, at present the most advanced 
and most efficient system to compute stable models of DATALOG$^\neg$ 
programs, the well-founded semantics can be used as a powerful lookahead 
mechanism. 

Despite the importance of the well-founded semantics, the question of how
fast it can be computed has not attracted significant attention. Van Gelder 
\cite{vg89} described the so called {\em alternating-fixpoint} algorithm. 
Van Gelder's algorithm runs in time $O(|\At(P)|\times size(P))$, 
where $\At(P)$ is the set of atoms occurring in a logic program $P$,
$|\At(P)|$ denotes the cardinality of $\At(P)$, and $size(P)$ is the size of 
$P$ (the total number of atom occurrences in $P$). Improving on this 
algorithm turned out to be difficult. The first progress was obtained in
\cite{bsf95}. The algorithm described there, when restricted to programs
whose rules contain at most two positive occurrences of atoms in their
bodies, runs in time $O(|\At(P)|^{4/3}|P|^{2/3} + size(P))$, where 
$|P|$ is the number of rules in $P$. For programs whose rules have no 
more than one positive atom in the body a better estimate of 
$O(|\At(P)|^{3/2}|P|^{1/2} + size(P))$ was obtained. For some classes 
of programs this is an asymptotically better estimate than the
$O(|\At(P)|\times size(P))$ estimate that holds for  
the algorithm by Van Gelder. 

A different approach to computing the well-founded model was proposed 
in \cite{zbf97,bdfz99}. It is based on the notion of a program 
transformation \cite{bd98}. The authors describe there several 
transformations that can be implemented in linear time and that
simplify a program while (essentially) preserving the well-founded 
semantics. These transformations are: the positive reduction, success, 
negative reduction, and failure (PSNF transformations, for short). 
They allow one to compute in linear time the Kripke-Kleene 
semantics \cite{fi85} of the program. 
To compute the well-founded semantics one also 
needs to detect the so-called {\em positive loops}. The complexity of this 
task dominates the asymptotic complexity of the well-founded semantics 
computation. No improved algorithms for the positive-loop detection
are offered in \cite{bdfz99} so the worst-case asymptotic complexity of 
the algorithm presented there remains the same as that of 
the alternating-fixpoint method. However, due to the use of PSNF 
transformations, that simplify 
the program, the algorithm based on program transformations may 
in practice run faster. In contrast to the approach studied in 
\cite{bdfz99}, we focus here on the positive-loop detection task.

The alternating-fixpoint algorithm works by successively improving lower
approximations $T$ and $F$ to the sets of atoms that are true and false
(under the well-founded semantics), respectively. The algorithm
starts with $T=\emptyset$. Using this estimate, it computes the first
estimate for $F$. Next, using this estimate, in turn, it computes 
a better estimate for $T$. The algorithm continues until further improvements
are not possible. It returns the final sets $T$ and $F$ as the well-founded
semantics. A dual version of the
alternating-fixpoint algorithm, starting with $F=\emptyset$ and then
alternatingly computing approximations to $T$ and $F$, is also possible. 
The most time-consuming part of this algorithm is in
computing estimates to the set of atoms that are false (in this part, in 
particular, positive loops are detected). In the Van Gelder algorithm, 
the best possible approximation (given the current estimate for $T$) is 
always computed by using a bottom-up approach. 

In this paper we focus on the problem of detecting positive loops and 
computing new false atoms. We restrict our attention to the class of
programs that have at most one positive atom in the body. We denote 
this class of programs by ${\cal LP}_1$. We show that for programs from
${\cal LP}_1$, false atoms can be computed by means of a top-down 
approach by finding atoms that do not have a proof. Moreover, we show 
that it is not necessary to find {\em all} atoms that can be established 
to be false at a given stage. Finding a proper subset (as long as it 
is not empty) is also sufficient and results in a correct algorithm.
We apply these techniques to design a version of an alternating-fixpoint 
algorithm computing the well-founded semantics of programs from
the class ${\cal LP}_1$. We demonstrate that the resulting algorithm is 
asymptotically better than the original alternating-fixpoint algorithm 
by Van Gelder. Specifically, we show that our algorithm runs in 
time $O(|\At(P)|^2 + size(P))$. Thus, for programs with $size(P)\geq 
|\At(P)|^2$, our algorithm runs in linear time and is asymptotically 
optimal! It is also easy to see that when $|P| > |\At(P)|$, the 
asymptotic estimate of the running time of our algorithm is better 
than that of algorithms by Van Gelder \cite{vg89} and Berman et 
al. \cite{bsf95}. 

As mentioned above, our approach is restricted to the class ${\cal LP}_1$.
Applicability of our method can, however, be slightly extended. Let us 
denote by ${\cal LP}_1^+$ the class of these logic programs that, after 
simplifying by means of PSNF transformations (or, equivalently, with 
respect to the Kripke-Kleene semantics) fall into the class ${\cal LP}_1$. 
Since PSNF transformations (the Kripke-Kleene semantics) can be computed in 
linear time, the asymptotic estimate of the running time of our method 
extends to all programs in the class ${\cal LP}_1^+$.

The paper is organized as follows. In the next section we provide a
brief review of the key notions and terminology. In Section \ref{algs}
we describe several modifications to the original Van Gelder algorithm,
we show their correctness and estimate their running time. The ultimate
effect of our considerations there is a general template for an algorithm 
to compute the well-founded semantics. Any algorithm computing some (not
necessarily all) atoms that can be established as false given a current 
estimate to the well-founded can be used with it. One such algorithm, 
for programs from the class ${\cal LP}_1$, is described and
analyzed in Section \ref{cluster}. It constitutes the main contribution
of the paper and yields a new, currently asymptotically most efficient 
algorithm for computing the well-founded semantics for programs in
${\cal LP}_1$. The last section contains conclusions.

\section{Preliminaries}
\label{prel}

We start by reviewing basic concepts and notation related to logic
programs and the well-founded semantics, as well as some simple auxiliary
results. In the paper we consider the propositional case only. 

Let $P$ be a normal logic program. By $\At(P)$ we denote the set of atoms 
occurring in $P$. Let $M\subseteq \At(P)$ (throughout the paper we often
drop a reference to $P$ from our notation, whenever there is no danger of
ambiguity). By $P_M$ we denote the program obtained from $P$ by removing 
all rules whose bodies contain negated literals of the form $\n(a)$, 
where $a\in M$. Further, by $P^h$ we denote the program obtained from $P$ 
by removing from the bodies of its rules {\em all} negative literals. 
Clearly, the program $(P_M)^h$ coincides with the 
{\em Gelfond-Lifschitz} reduct of $P$ with respect to $M$ (throughout
the paper, we write $P_M^h$ for $(P_M)^h$, to simplify notation). The 
{\em Gelfond-Lifschitz} operator on the algebra of all subsets of
$\At$, $GL$ (following our convention, we omit the reference to $P$ from 
the notation), is defined by
\[
GL(M) = LM(P_M^h),
\]
where $LM(Q)$ stands for a least model of a Horn program $Q$.

We now present characterizations of the well-founded semantics. We
phrase them in the language of operators and their fixpoints. All
operators considered here are defined on the algebra of subsets of
$\At(P)$. We denote a least fixpoint (if it exists) of an operator
$O$ by $\lfp(O)$.

It is well known that $GL$ is antimonotone. Consequently, $GL^2=GL
\circ GL$ is monotone and has a least fixpoint. The set of atoms that
are true with respect to the well-founded semantics of a program $P$,
denoted by $T_{\wfs}$, is precisely the least fixpoint of the operator
$GL^2$, that is, $T_{\wfs}= \lfp(GL^2)$ \cite{vg89,fi99}. The set of atoms
that are false with respect to the well-founded semantics of a program $P$,
denoted by $F_{\wfs}$, is given by $\overline{GL(T_{\wfs})}$ (throughout
the paper, $\overline{X}$ denotes the complement of a set $X$ with
respect to $\At(P)$). 

One can define a dual operator to $GL^2$ by
\[
A(M) = \overline{GL(GL(\overline{M}))}.
\]
It is easy to see that $A$ is monotone and that its least fixpoint
is $F_{\wfs}$. Thus, $F_{\wfs}=\lfp(A)$ and $T_{\wfs}=GL(\overline{F_{\wfs}})$. 
 
We close this section by discussing ways to compute $GL(M)$ for a given
finite propositional logic program $P$ and a set of atoms $M\subseteq 
\At(P)$. A straightforward approach is to compute the Gelfond-Lifschitz
reduct $P_M^h$ and then to compute its least model. The resulting
algorithm is asymptotically optimal as it runs in time linear in the
size of the program. However, in this paper we will use a different 
approach, more appropriate for the computation of the well-founded
semantics. Let $P$ be a logic program with negation. 
We define $\At^-(P) = \{\n(a) \colon a\in \At(P)\}$. For every set 
$M\subseteq \At(P)\cup \At^-(P)$, we define $true(M)= M\cap \At(P)$. 
If we interpret literals of $\At^-(P)$ as new {\em atoms}, then for 
every set $M\subseteq \At(P)$, the program $P\cup \n(M)$ can be viewed 
as a Horn program. Thus, it has a least model. It is easy to see that 
\[
GL_P(M)= true(LM(P\cup \n(\overline{M}))).
\]
Here, $P$ appearing at the left-hand side of the equation stands for 
the original logic program, while $P$ appearing at the right-hand side
of the equation stands for the same program but interpreted as a Horn 
program. Thus, using the algorithm of Dowling and Gallier \cite{dg84}, 
the Gelfond-Lifschitz reduct can be computed in time
$O(size(P)+|M|)=O(size(P))$ (since $M\subseteq \At(P)$,
$|M|=O(size(P))$).

\section{Algorithms}
\label{algs}

The departure point for our discussion of algorithms to compute the
well-founded semantics is the {\em alternating-fixpoint} algorithm of
Van Gelder \cite{vg89}. Using the terminology introduced in the previous 
section it can be formulated as follows.

\begin{tabbing}
\quad\=\quad\=\quad\=\quad\=\quad\=\quad\=\quad\=\quad\=\quad\=\quad\=\quad\=\\
\bf Algorithm 1 (Van Gelder)\\
\>\>$F:=\emptyset$;\\
\>\>{\bf repeat}\\
\>\>\>$T:=true(LM(P\cup \n(F))$; (* or equivalently:
                         $T:=GL(\overline{F})$; *)\\
\>\>\>$F:= \overline{LM(P_T^h)}$; (* or equivalently:
                       $\overline{GL(T)}$; *)\\
\>\>{\bf until} no change in $F$;\\
\>\>{\bf return} $T$ and $F$.
\end{tabbing}

Let $F'$ and $F''$ be the values of the set $F$ just before and just after 
an iteration of the {\bf repeat} loop in Algorithm 1. Clearly,
\[
F'' = \overline{GL(GL(\overline{F'}))} = A(F').
\]
Thus, after iteration $i$ of the {\bf repeat} loop, $F=
A^i(\emptyset)$. Consequently, it follows from our earlier remarks that
when Algorithm 1 terminates, the set $F$ that is returned satisfies 
$F=F_{\wfs}$. Since there is no change in $F$ in the last iteration, 
when the algorithm terminates, we have $T=T_{\wfs}$. That is, Algorithm 1 
is correct. 

We will now modify Algorithm 1. The basis for Algorithm 1 is the
operator $A$. This operator is not {\em progressive}. That is, $M$ is not 
necessarily a subset of $A(M)$. We will now introduce a related 
progressive operator, say $B$, and show that it can be used to replace $A$. 
Let $P$ be a logic program and let $T$ and $F$ be two subsets of $\At(P)$. 
By $P_{F,T}$ we denote the program obtained from $P$ by removing
\begin{enumerate}
\item all rules whose heads are in $F$
\item all rules whose bodies contain a positive occurrence of an atom
from $F$ 
\item all rules whose bodies contain a negated literal of the form
$\n(a)$, where $a\in T$.
\end{enumerate}
Clearly, $P_{F,T}\subseteq P_T$.

We define an operator $B(F)$ as follows:
\[
B(F) = \overline{LM(P_{F,T}^h)},
\]
where $T=GL(\overline{F})$ and $P_{F,T}^h$ abbreviates $(P_{F,T})^h$.
The following result gathers key properties of the operator $B$.

\begin{theorem}\label{l1a}
Let $P$ be a normal logic program. Then:
\begin{enumerate}
\item \label{l11} $B$ is monotone
\item \label{l12} For every $F\subseteq \At(P)$, $A(F)\subseteq B(F)$
\item \label{l13} For every $F\subseteq F_{\wfs}$, $B(F)\subseteq
F_{\wfs}$
\item \label{l15} $\lfp(B) = F_{\wfs}$
\item \label{l14} For every $F\subseteq \At(P)$,
$B(F)=F\cup(\overline{F} \setminus LM(P_{F,T}^h))$, where
$T=GL(\overline{F})$.
\end{enumerate}
\end{theorem}
Proof: (\ref{l11}) Assume that $F_1\subseteq F_2$. Set
$T_i=GL(\overline{F_i})$, $i=1,2$. Clearly, $\overline{F_2}\subseteq
\overline{F_1}$ and, by antimonotonicity of $GL$, $T_1\subseteq T_2$.   
By the definition of $P_{F,T}$, $P_{F_2,T_2}\subseteq P_{F_1,T_1}$.
Consequently, $LM(P_{F_2,T_2}^h) \subseteq LM(P_{F_1,T_1}^h)$ and, so,
$B(F_1)\subseteq B(F_2)$.\\
\noindent
(\ref{l12}) Let $T=GL(\overline{F})$. Clearly, $P_{F,T}\subseteq P_T$.
Thus, $A(F) = \overline{LM(P_T^h)} \subseteq \overline{LM(P_{F,T}^h)} =
B(F)$.\\
\noindent
(\ref{l13}) We have, $LM(P_{T_\wfs}^h) = \overline{F_\wfs}$. It follows
that removing from $P_{T_\wfs}^h$ rules with heads in $F_\wfs$ and those
that contain an atom from $F_\wfs$ in their bodies does not
change the least model. That is, 
\[
LM(P_{F_\wfs,T_\wfs}^h) = LM(P_{T_\wfs}^h).
\]
Since, $T_\wfs = GL(\overline{F_\wfs})$,
$B(F_\wfs) = \overline{LM(P_{F_\wfs,T_\wfs}^h)}$. 
Let $F\subseteq F_\wfs$. Then, by (\ref{l11}), $B(F) \subseteq B(F_\wfs)$.
Thus, we have
\[
B(F) \subseteq B(F_\wfs) = \overline{LM(P_{F_\wfs,T_\wfs}^h)} =
\overline{LM(P_{T_\wfs}^h)} = {F_\wfs}.
\]
\noindent
(\ref{l15}) The least fixpoint of $B$ is given by $\lfp(B)=
\bigcup B^i(\emptyset)$. By (\ref{l13}), $\lfp(B)\subseteq
F_{\wfs}$. On the other hand, by (\ref{l11}) and (\ref{l12}),
$A^i(\emptyset) \subseteq B^i(\emptyset)$. Thus, $F_{\wfs} =\lfp(A)
\subseteq \lfp(B)$. It follows that $\lfp(B)=F_\wfs$.

\noindent
(\ref{l14}) Let $T=GL(\overline{F})$. Since $P_{F,T}$ has no rules with 
head in $F$, $LM(P_{F,T}^h) \subseteq \overline{F}$ and, consequently,
$F\subseteq B(F)$. Thus, the assertion follows.  \hfill$\Box$

Theorem \ref{l1a} allows us to prove the correctness of the following 
modification of Algorithm 1.

\begin{tabbing}
\quad\=\quad\=\quad\=\quad\=\quad\=\quad\=\quad\=\quad\=\quad\=\quad\=\quad\=\\
\bf Algorithm 2\\
\>\>$F:=\emptyset$;\\
\>\>{\bf repeat}\\
\>\>\>$T:=true(LM(P\cup \n(F))$;\\
\>\>\>$\Delta F:= \overline{F}\setminus LM(P_{F,T}^h)$;\\
\>\>\>$F:= F\cup \Delta F$;\\
\>\>{\bf until} no change in $F$;\\
\>\>{\bf return} $T$ and $F$.
\end{tabbing}

By Theorem \ref{l1a}, each iteration of the {\bf repeat} loop computes
$B(F)$ as the new value for the set $F$. More formally, the set
$F$ just after iteration $i$, satisfies $F= B^i(\emptyset)$. Thus, when
the algorithm terminates, the set $F$ that is returned is the least 
fixpoint of $B$. Consequently, by Theorem \ref{l1a}(\ref{l15}), 
Algorithm 2 is correct.
 
We will now modify Algorithm 2 to obtain a general template for an
alternating-fixpoint algorithm to compute the well-founded semantics. The 
key idea is to observe that it is enough to compute a
subset of $\Delta F$ in each iteration and the algorithm remains
correct. 

Let us assume that for some operator $\Delta_w$ defined for pairs
$(F,Q)$, where $F \subseteq \At(P)$ and $Q$ is a Horn program such that
$\At(Q)\subseteq \overline{F}$ (the complement is, as always, evaluated
with respect to $\At(P)$), we have:
\begin{description}
\item[\ (W1)] $\Delta_w(F,Q) \subseteq \overline{F} \setminus LM(Q)$
\item[\ (W2)] $\Delta_w(F,Q)=\emptyset$ if and only if $\overline{F}
\setminus LM(Q) =\emptyset$.
\end{description}

Let $F\subseteq\At(P)$. By the definition of $P_{F,T}$, $\At(P_{F,T}^h)
\subseteq \overline{F}$. Thus, we define $B_w(F) = 
F\cup \Delta_w(F,P_{F,T}^h)$, where $T=true(LM(P\cup \n(F)))$. 
It is clear that for every
$F\subseteq \At(P)$, $F\subseteq B_w(F) \subseteq B(F)$, the latter
inclusion follows from Theorem \ref{l1a}(\ref{l14}) and (W1). Consequently, 
for every $i$,
\[
B^i_w(\emptyset) \subseteq B^{i}(\emptyset).
\]
It follows that $B^i_w(\emptyset) \subseteq \lfp(B)=F_{\wfs}$. It also
follows that there is the first $i$ such that
$B^i_w(\emptyset)=B^{i+1}_w(\emptyset)$. Let us denote this set
$B^i_w(\emptyset)$ by $F_0$. Then $F_0\subseteq F_{\wfs}$. In the same time,
by condition (W2), $B(F_0) =F_0$. Since $F_{\wfs}$ is the least fixpoint of
$B$, $F_{\wfs}\subseteq F_0$. It follows that a modification of 
Algorithm 2 in which line 
\[
\Delta F:= \overline{F}\setminus LM(P_{F,T}^h);
\]
is replaced by
\[
\Delta F:= \Delta_w(F,P_{F,T}^h);
\]
correctly computes the well-founded semantics of a program $P$.
Thus, we obtain the following algorithm for computing the well-founded
semantics.

\begin{tabbing}
\quad\=\quad\=\quad\=\quad\=\quad\=\quad\=\quad\=\quad\=\quad\=\quad\=\quad\=\\
\bf Algorithm 3\\
\>\>$F:=\emptyset$;\\
\>\>{\bf repeat}\\
\>\>\>$T:=true(LM(P\cup \n(F))$;\\
\>\>\>$\Delta F:= \Delta_w(F,P_{F,T}^h)$;\\
\>\>\>$F:= F\cup \Delta F$;\\
\>\>{\bf until} no change in $F$;\\
\>\>{\bf return} $T$ and $F$.
\end{tabbing}

We will now refine Algorithm 3. Specifically, we will show that the 
sets $T$ and $F$ can be computed incrementally.

Let $R$ be a Horn program. We define the {\em residual} program of $R$,
$res(R)$, to be the Horn program obtained from $R$ by removing all rules
of $R$ with the head in $LM(R)$ and by removing from the bodies of the
remaining rules those elements that are in $LM(R)$. We have the
following technical result. 

\begin{lemma}\label{l0}
Let $R$ be a Horn program and let $M$ be a set of atoms such that
$M\cap head(R)=\emptyset$. Then $LM(R\cup M)= LM(R)\cup LM(res(R)\cup
M)$. \hfill $\Box$
\end{lemma}

Lemma \ref{l0} implies that (we treat here negated literals as new atoms
and $P$ as Horn program over the extended alphabet)
\[
LM(P\cup \n(F\cup \Delta F)) = 
LM(P\cup \n(F)) \cup LM(res(P\cup \n(F))\cup \n(\Delta F)).
\]
Thus, if the set $F$ is expanded by new elements from $\Delta F$, then 
the new set $T$ can be computed by increasing the old set $T$ by 
$\Delta T= true(LM(res(P\cup \n(F))\cup \n(\Delta F)))$. Important thing 
to note 
is that the increment $\Delta T$ can be computed on the basis of 
the residual program and the increment $\Delta F$. Similarly, we have
\[
P_{F\cup \Delta F,T\cup \Delta T} = (P_{F,T})_{\Delta F,\Delta T}.
\]
Thus, computing $P_{F,T}$ can also be done incrementally on the basis of
the program considered in the previous iteration by taking into account
most recently computed increments $\Delta F$ and $\Delta T$.

This discussion implies that Algorithm 3 can be equivalently restated 
as follows:

\begin{tabbing}
\quad\=\quad\=\quad\=\quad\=\quad\=\quad\=\quad\=\quad\=\quad\=\quad\=\quad\=\\
\bf Algorithm 3\\
1\>\>$T:=F:=\Delta T:=\Delta F:= \emptyset$;\\
2\>\>$R:=P$; (*$R$ will be treated as a Horn program *)\\
3\>\>$Q:=P$;\\
4\>\>{\bf repeat}\\
5\>\>\>$\Delta T:=true(LM(R\cup \n(\Delta F))$;\\
6\>\>\>$R:= res(R\cup \n(\Delta F))$;\\
7\>\>\>$T:=T\cup \Delta T$;\\
8\>\>\>$Q:=Q_{\Delta F,\Delta T}$;\\
9\>\>\>$\Delta F := \Delta_w(F,Q^h)$;\\
10\>\>\>$F:=F\cup \Delta F$;\\
11\>\>{\bf until} no change in $F$;\\
12\>\>{\bf return} $T$ and $F$.
\end{tabbing}

We will now estimate the running time of Algorithm 3. Clearly line 1
requires constant time. Setting up appropriate data structures for
programs $R$ and $Q$ (lines 2 and 3) takes $O(size(P))$ steps.
In each iteration, $\Delta T$ is computed and the current program $R$ 
is replaced by the program $\res(R\cup\n(\Delta F))$ (lines 5 and 6). 
By modifying the algorithm from \cite{dg84} and assuming that $R$
is already stored in the memory (it is
avaliable either as the result of the initialization in the case of the
first iteration or as a result of the computation in the previous
iteration), both tasks can be accomplished in $O(size(R^o)+|\Delta F| 
- size(R^n))$ steps. Here $R^o$ denotes the old version of $R$ and $R^n$ 
denotes the new version of $R$. Consequently, the total time needed for 
lines 5 and 6 over all iterations is given by $O(size(P)+|\At(P)|-
size(R^t)) = O(size(P))$ (where $R^t$ is the program $R$, when the 
algorithm terminates). The time needed for all lines 7 is proportional 
to the number of iterations and is $O(|\At(P)|)=O(size(P))$. 

Given a logic program $Q$ and sets of atoms $\Delta T$ and $\Delta F$,
it takes $O(size(Q) -size(Q_{\Delta F,\Delta T})+|\Delta T|+|\Delta F|)$ 
steps to compute the program $Q_{\Delta F,\Delta T}$ in line 8. We 
assume here that 
$Q$ is already in the memory as a result of the initialization in the 
case of the first iteration, or as the result of the computation in 
the previous iteration, otherwise. It follows that the total time over 
all iterations needed to execute line 8 is $O(size(P)+|\At(P)|)=
O(size(P))$.  

Thus, we obtain that the running time of Algorithm 3 is
given by $O(size(P)+m)$, where $m$ is the total time needed to compute
$\Delta_w(F,Q^h)$ over all iterations of the algorithm.

In the standard (Van Gelder's) implementation of Algorithm 3, we compute
the whole set $\overline{F}\setminus LM(Q^h)$ as $\Delta_w(F,Q^h)$. In
addition, computation is performed in a bottom-up fashion. That is, 
we first compute the least model of $Q^h$ and then 
its complement with respect to $\overline{F}$. Such approach requires
$O(size(Q^h))=O(size(P))$ steps per iteration to execute line 9 and 
leads to $O(|\At(P)|\times size(P))$ running-time estimate for the 
alternating-fixpoint algorithm.

\section{Procedure $\Delta_w$}
\label{cluster}

In this section we will focus on the class of programs, ${\cal LP}_1$,
that is, programs whose rules have no more than one positive atom in
their bodies. We assume that we have a procedure $\false$ that, given 
a Horn program $Q
\in {\cal LP}_1$, returns a subset of the set $\At(Q)\setminus LM(Q)$. 
We also assume that $\false$ returns the empty set {\em if and only if} 
$\At(Q)=LM(Q)$. 
For every pair $(F,Q)$, where $F\subseteq \At(P)$ and $Q$ is a Horn program 
such that $\At(Q)\subseteq \overline{F}$, we define
\[
\Delta_w(F,Q) = \false(Q).
\]
It is easy to see that this operator $\Delta_w(F,Q)$ satisfies 
conditions (W1) and (W2). Consequently, it can be used in Algorithm 3.
Clearly, the procedure $\Delta_w$ and its computational properties are
determined by the procedure $\false$. In the remainder of the paper, we 
will describe a particular implementation of the procedure $\false$ and 
estimate its running time. We will use this estimate to obtain a bound on 
the running time of the resulting version of Algorithm 3. 

A straightforward way to compute the least model of $Q$ and so, to 
find $At(Q)\setminus LM(Q)$, is "bottom-up". That is, we start 
with atoms which are heads of rules with the empty bodies and use the 
rules of $Q$ to compute all atoms in $LM(Q)$ by iterating the van
Emden-Kowalski operator. An efficient implementation of the process is
provided by the Dowling-Gallier algorithm \cite{dg84}.

The approach we follow here in the  procedure $\false$ is "top-down" 
and gives us, in general, only a part of the set $At(Q)\setminus LM(Q)$. More 
precisely, for an atom $a$ we proceed ``backwards'' attempting to 
construct a proof or to demonstrate that no proof exists. In the process, 
we either go back to an atom that is the head of a rule with empty body 
or we show that no proof exists. In the former case, $a\in LM(Q)$.
In the latter one, none of the atoms considered while searching for a 
proof of $a$ are in $LM(Q)$ (because $Q\in{\cal LP}_1$ and each rule has
at most one antecedent). The problem is that we may 
find an atom $a$ that does not have a proof only {\em after} we look at 
all other atoms first. Thus, in the worst case, finding one new false 
atom may require time that is proportional to the size of $Q$. 

To improve the time performance, we look for proofs simultaneously for 
all atoms and grow the proofs ``backwards'' in a carefully controlled 
way. Namely, we never let one search to get too much ahead of the other
searches. This controlled way of looking for proofs 
is the key idea of our approach and leads to a better performance.
We will now provide an informal description of the procedure $\false$
followed later by a formal specification and an example.

In the procedure, we make use of a {\em new} atom, say $s$, different 
from all atoms occurring in $Q$. Further, we denote by $head(r)$ the 
atom in the head of a rule $r\in Q$ and by $tail(r)$ the atom which is 
either the unique positive atom in the body of $r$, if such an atom exists, 
or $s$ otherwise. We call an atom $a\in At(Q)$ {\it accessible} if there 
are rules $r_1,\ldots,r_k$ in $Q$ such that $tail(r_{i+1})=head(r_i)$, 
for $i=1,\ldots,k-1$, $tail (r_1)=s$ and $head(r_k)=a$. Clearly, the least 
model $LM(Q)$ of $Q$ is precisely the set of all accessible atoms.

In each step of the algorithm, the set of atoms from $\At(Q)$ is
partitioned into {\em potentially false sets} or {\em pf-sets}, for
short. We say that a set $v\subseteq At(Q)$ is a {\em pf-set} if
for each pair of {\em distinct}
atoms $a,b\in v$ there are rules $r_1,\ldots,r_k$ 
in $Q$ such that $tail(r_{i+1})= head(r_i)\in v$, for $i=1,\ldots,k-1$, 
$tail (r_1)=b$ and 
$head(r_k)=a$. It is clear that if $v$ is a pf-set then either all its 
elements are accessible (belong to the least model of $Q$) or none of them
does (they are all false). Clearly, singleton sets consisting of individual 
atoms in $At(Q)$ are pf-sets. In the algorithm, with each pf-set we 
maintain its cardinality.

Current information about the state of all top-down searches and about
the dependencies among atoms, that were discovered so far, is maintained in
a directed graph $\cal G$. The vertex set of this graph, say ${\cal S}$,
consists of $\{s\}$ and of a family of pf-sets forming a partition of 
the set $\At(Q)$. The edges of $\cal G$ are specified by a {\em
partial} function $\pred:{\cal S} \rightarrow {\cal S}$. We write
$\pred(v)={\bf undefined}$ if $\pred$ is undefined for $v$. 
Thus, the set of edges of $\cal G$ is given by $\{(\pred(v),v)\colon 
\pred(v) \not={\bf undefined}\}$. Since $\pred$ is a partial function, it is 
easy to see that the connected components of the graph $\cal G$ are 
unicyclic graphs or trees rooted in those vertices $v$ for which $\pred(v)$ 
is undefined. Throughout the algorithm we always have
$\pred(\{s\})={\bf undefined}$. Thus, the connected component of 
$\cal G$ containing $\{s\}$ is always a tree and $\{s\}$ is its root.

If $w$ and $v$ are two different pf-sets,
the existence of the edge $(w,v)$ in $\cal G$ means that we 
have already discovered
a rule in the original program whose head is in $v$ and whose tail is in 
$w$. Thus, if vertices in $w$ are accessible, then so are the vertices in 
$v$. A pf-set that is the root of a tree forming a component of 
$\cal G$ is called an {\em active} pf-set. If $v$ is an active pf-set then 
no rule $r$ with $head(r)\in v$ and $tail(r)\not\in v$ has been detected 
so far. Thus, $v$ is a candidate for a set of atoms which does not intersect 
the least model of $Q$. Let us note that even though $\{s\}$ is a root of 
a tree in $\cal G$ it is never active as it is not a pf-set in the first 
place.

We let active pf-sets grow by gluing them with other pf-sets. However, 
we allow to grow only these active pf-sets whose cardinalities 
are the least. In each iteration of the algorithm the value of the 
variable $size$ is a lower bound for the cardinalities of active pf-sets.
To grow an active pf-set $v$, we look for rules with heads in $v$ and
with tails in pf-sets {\em other} than $v$ (not necessarily active) or 
in $\{s\}$. The 
dependencies between pf-sets discovered in this way are represented
as new directed edges in $\cal G$. Pf-sets that appear in the same cycle
are glued together (in the procedure {\em cycle}). Since $\{s\}$ is not
an active pf-set, it never becomes an element of a cycle in $\cal G$. 

If, when attempting to grow 
a pf-set $v$ we discover a rule with head in $v$ and with the tail
in a vertex of the tree of $\cal G$ rooted in $\{s\}$,
then $v$ is from now on ignored (all its vertices
belong to the least model of $Q$). Indeed, $v$ gets connected to a tree
of $\cal G$ rooted in $\{s\}$. Consequently, it cannot become a member
of a cycle in $\cal G$ in the future and is never again considered by
the procedure {\em cycle}.

The main loop (lines 6-23) of the algorithm $\false$ below starts
by incrementing $size$ followed by a call to the procedure $cycle({\cal 
S}, \pred,size,L)$. This procedure scans the graph $\cal G$ and 
identifies all its cycles. It then modifies $\cal G$ by considering each 
cycle and by gluing its pf-sets into a single pf-set. To this end, it
modifies the vertex set $\cal S$ of $\cal G$ and the function $\pred$ 
defining the edges of $\cal G$. Each such new pf-set 
becomes the root of its tree in $\cal G$ and so, it becomes active. 
The procedure {\em cycle} computes the cardinality of each new active
pf-set. Finally, it creates a list $L$ so that it consists of active pf-sets 
of cardinality $size$. If no such set is found ($L$ is empty), we move on 
to the next iteration of the main loop and increment $size$ by 1. We
give a more detailed description of the procedure $cycle$ later in the
paper when we analyze the time complexity of our method.

For each active pf-set $v \in L$ we consider the tail of each rule with 
head in $v$ (lines 9-22). If there is a rule $r$ with $head(r)\in v$ and 
$tail(r)\not\in v$ then it is detected (line 15). The value $\pred(v)$ is 
set to this element in $\cal S$ that contains $tail(r)$ (it may be that
this set is $\{s\}$). We also set the variable $success$ to {\bf true}
(line 16). The pf-set $v$ stops to be active. We move on to the next 
active pf-set on $L$.

If such a rule $r$ does not exist then $success={\bf false}$ and $v$
is a set of cardinality $size$ consisting of atoms which are not in
the least model of $Q$. This set is returned by the procedure $\false$
(line 21). Hence, for an active pf-set considered in the loop
6-23, either we find a pf-set $\pred(v)\in {\cal
S}\setminus\{v\}$ (and we
have to consider the next pf-set on $L$) or $v$ is returned as a set 
of atoms which are not in the least model of $Q$
(and the procedure $\false$ terminates). Thus, the procedure $\false$ 
is completed if either a nonempty set $v$ of atoms which are not in 
the least model of $Q$ is found or, after some passes of the loop
6-23, the graph $\cal G$ has no active pf-sets. In the latter
case $\cal G$ is a tree with the root in $\{ s\}$. Thus, $At(Q)=LM(Q)$
and $v=\emptyset$ is returned (line 24).

In the procedure $\false$, as formally described below, an input program 
$Q$ is represented by lists $IN(a)$, $a\in At(Q)$, of all atoms $b$ such 
that $b$ is the body of some rule with the head $a$. If there is a rule 
with the head $a$ and empty body, we insert $s$ into the list $IN(a)$.

We also use an operation $next$ on lists and elements. Let $l$ be a list 
and $w$ be an element, either belonging to $l$ or having a special value 
{\bf undefined}. Then
\[
next(w,l)=\left\{ \begin{array}{ll}
{\rm the\ next\ element\ after}\ w\ {\rm in}\ l &{\rm \ \ \ if}\ w\in
l\\
{\rm the\ first\ element\ in}\ l &{\rm \ \ \ if}\ w\ {\rm is}\ {\bf
undefined}.
\end{array}
\right.
\]
The value {\bf undefined} should not be mixed with {\bf nil} which
indicates the end of a list.

Finally, we use a procedure $\findset(w,{\cal S})$ which, for an atom $w$ and 
a collection $\cal S$ of disjoint sets, one of which contains $w$, finds 
the name of the set in $\cal S$ containing $w$ (it follows from our
assumptions that such a set is unique). Elements of $\cal S$ are
maintained as linked lists. Each element on such a list has a pointer to
the head of the list. The head serves as the identifier for the list.
When the procedure $\findset(w,{\cal S})$ is called, it returns the head
of the list to which $w$ belongs.

\begin{tabbing}
\quad\=\quad\=\quad\=\quad\=\quad\=\quad\=\quad\=\quad\=\quad\=\quad\=\quad\=\\
1\>\>\bf procedure $\false(Q)$;\\
2\>\>\>${\cal S}:=\{\{ x\} : x\in At(Q)\} \cup \{ \{ s\}\}$;\\
3\>\>\>{\bf for} $v\in {\cal S}$ {\bf do} $\pred(v):={\bf undefined}$;\\
4\>\>\>{\bf for} $x\in At(Q)$ {\bf do} \{$w(x):={\bf undefined}$;\ \
$cardinality(x):= 1$\};\\
5\>\>\>$size:=0$;\\
6\>\>\>{\bf while} $size < |At(Q)|$ {\bf do}\\
7\>\>\>\>$\{size:=size+1$;\\
8\>\>\>\>$cycle({\cal S},\pred,size,L)$;\\ 
9\>\>\>\>{\bf for} all $v\in L$ {\bf do}\\ 
10\>\>\>\>\>$\{ success:=\bf false$;\\
11\>\>\>\>\>$u:=next(u,v)$;\\
12\>\>\>\>\>{\bf while} $u\not=$ {\bf nil and} {\bf not} $success$ {\bf do}\\
13\>\>\>\>\>\>$w(u):=next(w(u),IN(u))$;\\
14\>\>\>\>\>\>{\bf while} $w(u)\not=$ {\bf nil and} {\bf not} $success$ {\bf do}\\
15\>\>\>\>\>\>\>$\{${\bf if} $\findset(w(u),{\cal S})\not= v$\\
16\>\>\>\>\>\>\>\>{\bf then} $\{ success:={\bf true}$; $\pred(v):=\findset(w(u),{\cal S})\}$\\
17\>\>\>\>\>\>\>\>{\bf else} $w(u):=next(w(u),IN(u))$ \\
18\>\>\>\>\>\>{\bf end while} (14)$\}$;\\
19\>\>\>\>\>\>{\bf if} {\bf not} $success$ {\bf then} $u:=next(u,v)$\\
20\>\>\>\>\>{\bf end while} (12)$\}$;\\
21\>\>\>\>\>{\bf if} {\bf not} $success$ {\bf then return} $v$ \ \ \
(* the procedure terminates *)\\
22\>\>\>\>{\bf end for} (9)$\}$\\
23\>\>\>{\bf end while} (6)$\}$;\\
24\>\>\>\bf return $v=\emptyset$\\ 
25\>\>{\bf end} $\false$;\\
\end{tabbing}

We will now illustrate the operation of the algorithm. Let us consider
the following Horn logic program $Q$:
\[
\begin{array}{llllll}
a \leftarrow   &
b \leftarrow a &
a \leftarrow c &
c \leftarrow a &
a \leftarrow e &
d \leftarrow e \\

f \leftarrow d\ \ \ &
e \leftarrow f\ \ \ &
d \leftarrow f\ \ \ &
e \leftarrow g\ \ \ &
g \leftarrow j\ \ \ &
j \leftarrow g\ \ \ \\

i \leftarrow j &
j \leftarrow h &
k \leftarrow j &
k \leftarrow h &
h \leftarrow k &
\ 
\end{array}
\]

\begin{figure*}[htb]
\centerline{\psfig{figure=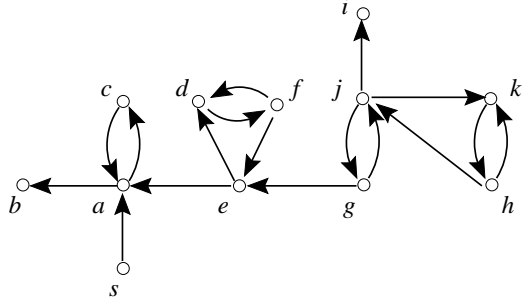}}
\caption{Graph $G^Q$ representing program $Q$.}
\label{rys1}
\end{figure*}

This program is represented as a graph, $G^Q$, in Fig. \ref{rys1}. The 
vertices of this graph correspond to the atoms of the program. In 
addition, $G^Q$ has an auxiliary vertex $s\notin \At(Q)$. An edge $(x,y)$, 
where $x,y\in \At(Q)$, represents the clause $y\leftarrow x$ from $Q$. 
An edge $(s,y)$, where $y\in \At(Q)$, represents the clause 
$y \leftarrow\ $. When illustrating the algorithm, we assume that atoms
from $\At(Q)$ (atoms $a, \ldots, k$ in our example) appear on the 
lists $IN(x)$, $x\in \At(Q)$, in the alphabetical order. We also assume 
that whenever $s$ belongs to a list $IN(x)$, it appears as the first atom 
on the list.

\begin{figure*}[htb]
\centerline{\psfig{figure=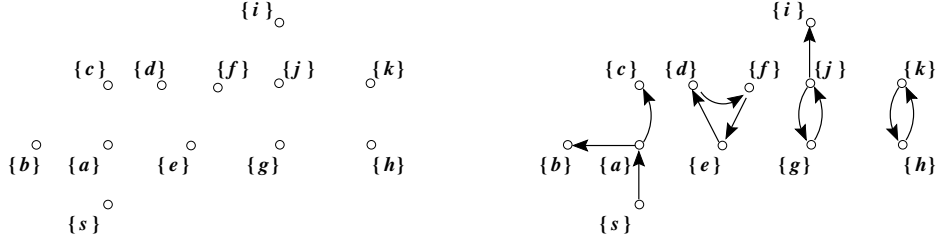}}
\caption{Graph $\cal G$ initially (left side) and after the first 
iteration of the loop 6-23 (on the right).}
\label{rys1a}
\end{figure*}

In the algorithm {\em false}, the current state of knowledge about the
possibility of proving an atom from $Q$ is represented by the graph $\cal
G$. Initially, $\cal G$ consists of isolated vertices. Indeed, line 3 of 
the algorithm sets $\pred(x)$ to undefined, for every vertex $x$ of $\cal S$
(see Fig. \ref{rys1a} (left)). All of the vertices of $\cal G$, except
for $\{s\}$ are active pf-sets. The procedure $cycle$ (line 8), 
called with $size=1$, 
puts all of them on the list $L$.

The algorithm considers next (line 9) all elements on the list $L$, that 
is, all vertices of $\cal G$ that are active pf-sets and have cardinality
equal to {\em size}. During the first 
iteration of the loop 6-23, $L$ consists of all vertices 
of $\cal G$, except for $\{s\}$ (that is, singleton sets $\{x\}$, where 
$x\in \At(Q) = 
V(G^Q)\setminus \{s\}$). For each vertex $v$ of $\cal G$ on $L$, 
the algorithm looks 
for a {\em back rule} for $v$, that is, a rule in $Q$ with the head in
$v$ and the tail in a pf-set other than $v$ or in $\{s\}$. In our
graphical representation of $Q$ by means of the graph $G^Q$, a back rule
for $v$ corresponds to an edge (referred to as a {\em back} edge) in $G^Q$ 
with the head in $v$ and the tail in a vertex of $\cal G$ other than 
$v$ (possibly in $\{s\}$). To find a back rule (edge) for $v$, all atoms 
$u$ of $Q$ (equivalently, all vertices $u$ of $G^Q$) that belong to $v$ 
are considered (the loop 12-20). For each such atom $u$, the algorithm 
searches for the first atom on the list $IN(u)$ that does not belong 
to $v$. Let us recall that $IN(u)$ is the list of
atoms that are the tails of rules with the head $u$ or, in the terms of
the graph $G^Q$, that are the tails of edges with the head $u$. 
If such an atom is found, together with $u$ 
it determines a back rule (edge) $r$ for $v$. The algorithm sets 
$\pred(v)$ to be equal to the pf-set containing the tail of $r$  
(line 16). That is, an edge from $\pred(v)$ to $v$ is added to $\cal G$. 
The algorithm moves then on to the next element of the list $L$. 

In our example, in the first iteration of the loop 6-23, 
a back rule is found for every element on $L$, that is, for every vertex 
of $\cal G$ other than $\{s\}$. For instance, for the vertex $\{d\}$,
the algorithm considers atoms on the list $IN(d) = (e,f)$ (let us recall
that atoms on lists $IN(x)$ are arranged alphabetically with the exception 
of the special atom $s$ which, if present on a list, is always its first 
element). The first atom on the list, $e$ does not belong to $\{d\}$.
Thus, it defines, together with $d$ a back rule for $\{d\}$,
$d\leftarrow e$. The resulting graph $\cal G$
is shown in Fig. \ref{rys1a} on the right. 

Let us note that when scanning the list $IN(d)$ in 
subsequent iterations the algorithm resumes the scan with the first atom
that has not been looked at yet (cf. the definition of the operation
{\em next}). Thus, the next time $d$ is considered as an element of an
active pf-set for which a back rule is searched for, the scan of $IN(d)$
will start with $f$. The same holds true for all lists $IN(x)$, $x\in
\At(Q)$. Consequently, each atom on each of these lists is considered
just once. Such an approach still guarantees that finding back rules
works correctly (that is, that they are found by the algorithm whenever
they exist). Indeed, when an atom on a list $IN(x)$ is considered, it
either defines a back rule with the head $x$ (and, thus, cannot define 
any new back rule with the head $x$ in the future) or it is in the same 
active pf-set as $x$ (and, thus, it neither defines a back rule now 
nor it will define it in the future, as it will remain in the same 
pf-set as $x$ till the algorithm terminates).

\begin{figure*}[htb]
\centerline{\psfig{figure=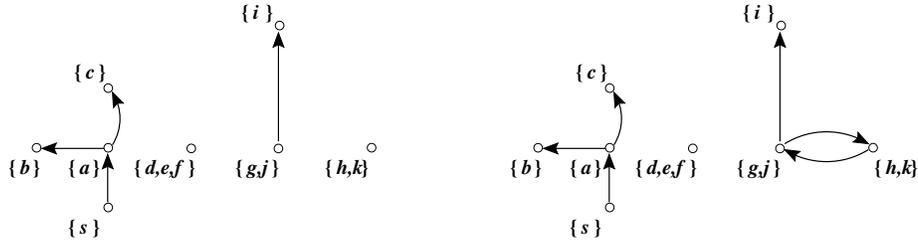}}
\caption{Graph $\cal G$ after the execution of the procedure 
{\em cycle} in the second iteration of the loop 6-23 (left) and after 
the second iteration of the loop 6-23 (right).}
\label{rys2a}
\end{figure*}

The second iteration of the loop 6-23 starts with 
the procedure {\em cycle} contracting each cycle in the graph 
$\cal G$ to a single vertex. The resulting graph is shown in 
Fig. \ref{rys2a} on the left. The procedure {\em cycle}
then creates a new list $L$. It consists of all active pf-sets of 
cardinality 2. In our case, $L$ contains $\{g,j\}$ and $\{h,k\}$ 
($\{d,e,f\}$ is also active but has cardinality 3).

Continuing with the second iteration, the algorithm next considers each 
vertex on $L$ (the loop 9-22) and looks for back rules. In this iteration, 
a back rule is found for each of the nodes on $L$ and the modified 
graph $\cal G$ is given in Fig. \ref{rys2a} on the right. 

\begin{figure*}[htb]
\centerline{\psfig{figure=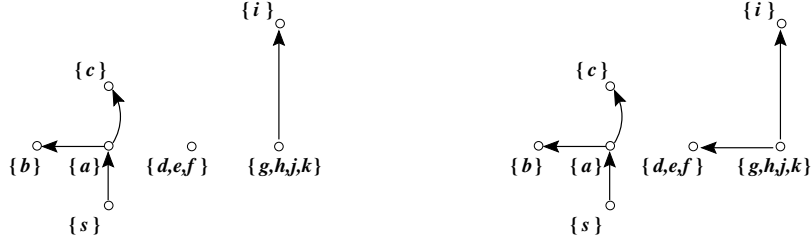}}
\caption{Graph $\cal G$ after the execution of the procedure
{\em cycle} in the third iteration of the loop 6-23 (left) and after
the third iteration of the loop 6-23 (right).}
\label{rys3a}
\end{figure*}

In the third iteration, the procedure {\em cycle} contracts the only
cycle in $\cal G$ to a single active pf-set of cardinality 4 
(Figure \ref{rys3a}, left side). It also creates a new list $L$.
This time it consists of active pf-sets of cardinality 3. There is just 
one such set - $\{d,e,f\}$. Subsequently, the algorithm {\em false} 
looks for a back rule for $\{d,e,f\}$. It starts by considering edges 
ending in $d$ (line 11; we assume that $v$ is represented by the list 
$(d,e,f)$). It scans the list $IN(d)$ starting at the first atom that
has not been inspected so far, that is, $f$. 
However, since $f$ belongs to the same pf-set as $d$, $f$ does not 
specify a back rule. Since there are no more atoms on the list $IN(d)$,
we move on to the next iteration of the loop 12-20 and consider atom 
$e$. We have $IN(e) = (f,g)$. Since $f$ was already considered (and
yielded a back rule for $\{e\}$) in the first iteration, we consider $g$.
Since $g \notin \{d,e,f\}$, it defines a back rule for $\{d,e,f\}$, 
$d \leftarrow g$.

The resulting graph $\cal G$ is shown in Figure \ref{rys3a} (on the 
right). It has no cycles. So, the only thing done by the procedure 
{\em cycle} in the iteration 4 is that it puts on $L$ active pf-sets of 
cardinality 4. There is just one such set in $\cal G$, $\{g,h,j,k\}$. 
The algorithm {\em false} looks for a back edge for $\{g,h,j,k\}$ and does
not find any. The variable {\em success} remains {\bf false}. The algorithm
returns $\{g,h,j,k\}$ and terminates (line 21). Let us note that this
set is a proper subset of the set $\At(Q)\setminus LM(Q)$.


The following theorem formally establishes two key properties of the
procedure $\false$. 

\begin{theorem}\label{t1}
\begin{enumerate}
\item The procedure $\false$ returns a set $v$ such that 
$v\subseteq At(Q)\setminus LM(Q)$.
\item $\false$ returns the empty set if and only if $At(Q)\setminus LM(Q)=\emptyset$.
\end {enumerate}
\end{theorem}
Proof:
(1) The statement is trivially true if {\it false} returns the empty 
set. Thus assume that the returned set $v\not= \emptyset$. It means that 
the value of the variable $success$ is {\bf false} after all passes of the 
loop 12-20 for some active pf-set $v$ in the list $L$. Thus 
every rule in $Q$ with the head in $v$ has been considered. 

Suppose there is a rule $r$ in $Q$ with $head(r)=u\in v$ and 
$tail(r)=b\not\in v$. This rule was considered by the procedure 
$\false$ when $u=head(r)$ was a member of some active pf-set, say $y$. 
Since larger pf-sets are obtained by gluing smaller ones, $y\subseteq v$. 
While $r$ was being considered, the value of $w(u)$ in the loop 14-18 
was $b$ and the value of $v$ was $y$. Consequently,
$\findset(b,{\cal S})\not= 
y$ in line 15 because $y\subseteq v$ and $b\not\in v$ so $b\not\in y$. 
Hence the value of $success$ was set to {\bf true} and $\pred(y)$ 
was defined to be, say, $z=\findset(b,{\cal S})$ in line 16. The pf-set $y$ 
stopped to be active. Recall that $v$ is active when the procedure 
stops. Hence $y$ had to be glued with other pf-sets to obtain $v$. 
This is, however, impossible because if $y$ were glued with some other 
pf-sets to form a larger pf-set $x$ then $\pred(y)=z\subseteq x$. 
Notice that $b\in z\subseteq x\subseteq v$. We have got a contradiction 
with $b\not\in v$. 

Hence, there are no rules $r$ in $Q$ with $head(r)\in v$ and $tail(r)\not
\in v$. Thus no atom in $v$ is accessible so $v\subseteq At(Q)\setminus LM(Q)$.\\
(2) Suppose {\it false} returns the empty set and consider the last 
pass of the loop 6-23, for $size=|At(Q)|$. If the list $L$ is 
empty then no vertex of $\cal G$ is an active pf-set. Hence, $\cal G$ is
a tree with the root $\{ s\}$. Thus all atoms in $At(Q)$ are accessible and 
consequently $LM(Q)=At(Q)$.

If the list $L$ is nonempty then it contains one pf-set $v=At(Q)$. The 
empty set is returned by the procedure $\false$ so the value of the variable 
$success$ in line 16  is {\bf true} for $v=At(Q)$. It means that for some 
rule $r$ in $Q$ with $head(r)=u$, $w(u)=tail(r)\not\in v=At(Q)$ so
$w(u)=s$. Hence, $u$ is accessible and, consequently, all atoms in $At(Q)$ 
are accessible. That is, we have $At(Q)\setminus LM(Q)=\emptyset$. 

The converse of the implication proved above follows immediately from the 
first part of the theorem. \hfill$\Box$

We shall now consider the procedure $cycle$ a little bit more carefully. 
The procedure can be informally written in the following form.

\begin{tabbing}
\quad\=\quad\=\quad\=\quad\=\quad\=\\
{\bf procedure} $cycle({\cal S},\pred,size,L)$\\
\smallskip
1.\>Initialize $L$ to empty.\\
2.\>Find all cycles $C_1,C_2,\ldots,C_p$ in the graph $\cal G$. Put ${\cal
C}=\{C_1,C_2,\ldots,C_p\}$.\\
3.\>For every cycle $C=\{v_1,\ldots,v_{q}\}$, $C\in {\cal C}$, do (i)-(iv).\\
\>(i)\>\>set $v_C:=v_1\cup \ldots\cup v_{q}$;\\
\>(ii)\>\>compute $cardinality(v_C)$ (sum up the cardinalities 
of all vertices in $C$);\\
\>(iii)\>\>update the set $\cal S$: set ${\cal S}:=({\cal S}-\{
v_1,\ldots,v_{q}\} )\cup \{ v_C\}$; (* $v_C$ becomes\\
\>\>\>an active pf-set *)\\
\>(iv)\>\>update the function $\pred$: for every $i=1,\ldots,q$, if 
$\pred(z)=v_i$ (for \\ 
\>\>\>some $z\in {\cal S}$) then $\pred(z):=v_C$;\\
4.\>For every vertex of $\cal G$ that is an active pf-set, if
$cardinality(v)=size$, insert $v$\\
\>into the list $L$.
\end{tabbing}

Since $\cal G$ is a directed graph whose connected components are either 
unicyclic graphs or trees, step 2 of the procedure $cycle$ can be 
implemented in ${O}(|{\cal S}|)$ time. Since pf-sets are represented as
linked lists, with each node on the list pointing to the head of the
list, step (i) can be implemented to take $O(|v_C|)$ steps. The time needed
for step (ii) is, clearly, $O(|C|)$. Each execution of step (iii) takes
also ${O}(|C|)$. Finally, 
the running time of each execution of step (iv) is $O(m_C)$, where
$m_C$ is the size of the connected component of the graph $\cal G$ containing 
$C$. Thus, an iteration of the loop 3 for a cycle $C\in {\cal C}$ takes 
$O(|C|+m_C+|v_C|)$. Clearly, $|C|\leq m_C$. Moreover, 
$\sum_{C\in{\cal C}}m_C \leq |{\cal S}| -1 \leq |\At(Q)|$ and 
$\sum_{C\in{\cal C}}|v_C| \leq |\At(Q)|$ (they are all
disjoint subsets of $\At(Q)$). Thus, the total time needed for the loop 3 is
$O(|\At(Q)|)$. It is easy to see that the time needed for the loop 4 is also
$O(|\At(Q)|)$. Consequently, the running time of the procedure $cycle$ is 
$O(|\At(Q)|)$.

We are now in a position to estimate the running time of the procedure
$\false$.

\begin{lemma}\label{l1}
If the procedure $\false(Q)$ returns a nonempty set $v$, then 
the running time of $\false$ is $O(|v|\times |\At(Q)|)$.
If $\false(Q)$ returns the empty set then its running time 
is $O(|\At(Q)|^2)$.
\end{lemma}
Proof: Let $|\At(Q)|=n$ and $|v|=k$. As we have already observed the 
procedure $cycle$ runs in time ${O}(n)$. It is not hard 
to see that, since we represent all sets occurring in the procedure $\false$ 
as linked lists, with each node on a list pointing to the head of the
list, the operations: $\findset$  and $next$ require a constant time.

First assume that the output $v$ of the procedure $\false$ is nonempty. Let 
us estimate the number of passes of the {\bf while} and {\bf for}
loops in the 
procedure. Clearly, the loop 6-23 is executed $k$ times. Hence 
the total running time of all calls of the procedure $cycle$ is ${O}(kn)$. 
The number of passes of the loop 9-22 is not larger than 
$|L_1|+|L_2|+\ldots+|L_k|$, where $L_i$ denotes the list $L$ in an
iteration $i$ of the loop. Since $L_i$ is a list of disjoint pf-sets of 
cardinality $i$, $|L_i|\leq n$, for each $i=1,2,\ldots,k$. 
Hence the number of passes of the loop 9-22 can be very roughly 
estimated by $kn$. The loop 12-20 is executed at most 
\[
\sum_{i=1}^k\sum_{v\in L_i}|v|\leq kn
\]
times. This inequality follows from the fact that the sets $v$ in 
the lists $L_i$ are disjoint subsets of atoms so $\sum_{v\in L_i}|v|\leq n$. 
The estimation of the number of passes of the loop 14-18 is a 
little bit more complicated. First notice that in each execution of the 
loop we check a rule of the program $Q$ and rules are checked only one time. 
The rules $r$ checked in the loop have either both the head and the tail in 
some pf-set $v\in {\cal S}$ or $head(r)\in v$ and $tail(r)$ is in some 
other pf-set $u\in {\cal S}$. In the latter case $\pred(v)$ is defined 
in line 16. The number of executions of line 16  is not larger than
the number of passes of the loop 9-22 so it is bounded by $kn$. When 
the procedure returns the output, the pf-sets have cardinalities not 
larger than $k$. 
Hence the number of rules with both the head and the tail in the same 
pf-set that has been checked before the procedure stops is not larger than 
\[
\sum_{u\in {\cal S}}|u|(|u|-1)\leq (k-1)\sum_{u\in {\cal S}}|u|\leq (k-1)n.
\]

Thus the number of passes of the loop 14-18 in the whole procedure 
$\false$ is less than $2kn$. It follows that if the output $v$ of $\false$ 
is nonempty then the running time of $\false$ is $O(|v|\times |\At(Q)|)$.

Now consider the case when the procedure $\false$ returns the empty set. 
Clearly the number of passes of the loop 6-23 is $n$ so it takes 
${O}(n^2)$ time for all executions of the procedure $cycle$. Since the 
rules are checked in the loop 14-18 only one time, the number 
of passes of this loop is not larger than the number $m$ of rules 
in $Q$. Obviously $m\leq n^2$ so the running time of $\false$ in this 
case is $O(|\At(Q)|^2)$. \hfill$\Box$

By Lemma \ref{l1} and considerations in Section 3 we get an estimation of 
the running time of Algorithm 3. 

\begin{theorem}\label{t2}
If $P$ is a program whose rules have at most one positive atom in the body 
then Algorithm 3 can be implemented so that its running time is 
$O(|At(P)|^2 + size(P))$. \hfill$\Box$
\end{theorem}

\section{Conclusions}
\label{conc}

The method for computing the well-founded semantics described 
in this paper is a refinement of the basic alternating-fixpoint
algorithm. The key idea is to use a top-down search when identifying
atoms that are false. Our method is designed to work with programs whose
rules have at most one positive atom in their bodies (class ${\cal
LP}_1$). Its running time is $O(|\At(P)|^2 + size(P))$ (where $P$ is 
an input program). Thus, our algorithm is an improvement over other 
known methods to compute the well-founded semantics for programs in the 
class ${\cal LP}_1$. Our algorithm runs in linear time for the class 
of programs $P\in{\cal LP}_1$ for which $size(P)\geq |\At(P)|^2$.
However, it is not a linear-time algorithm in general. It is an open
question whether a linear-time algorithm for computing the well-founded
semantics for programs in the class ${\cal LP}_1$ exists.

Our results extend to the class ${\cal LP}_1^+$. However, the extension 
is straightforward and the class ${\cal LP}_1^+$ is still rather narrow. 
Moreover, it is not specified syntactically (it is described by means 
of the Kripke-Kleene semantics). The question arises whether our top-down 
approach to positive-loop detection can be generalized to any class of 
programs significantly extending the class ${\cal LP}_1$ and possessing
a simple syntactic description.

Finally, let us note that the general problem of computing the well-founded
semantics still remains a challenge. No significant improvement over the
alternating-fixpoint algorithm of Van Gelder has been obtained for the
class of arbitrary finite propositional logic programs. 

\section*{Acknowledgments}
This research was supported by the NSF grants CDA-9502645 and IRI-9619233.

\end{document}